\documentclass[12pt]{article}
\def\semidirect{\;\rlap{$\supset$}{\times}\;}

\begin{document}
\title{Gauge theory of gravity: Electrically charged solutions within
  the metric--affine framework\thanks{Presented at the Workshop on
    ``Gauge Theories of Gravitation'', Jadwisin, Poland, 4 to
    10-Sept-1997}} \author{Friedrich W. Hehl and Jos\'e
  Socorro\thanks{Permanent address: Instituto de Fisica de la
    Universidad de Guanajuato, Apartado Postal E-143, CP. 37150,
    Le\'on, Guanajuato, M\'exico.}\\ Institute for Theoretical
  Physics, University of Cologne\\ D-50923 K\"oln, Germany} 

\maketitle

\begin{abstract} We find a 
  class of electrically charged exact solutions for a toy model of
  metric-affine gravity. Their metric is of the
  Pleba\'nski-Demi\-a\'nski type and their nonmetricity and torsion are
  represented by a triplet of covectors with dilation, shear, and spin
  charges. {\it file jadwisin4.tex, 1998-03-03}
\end{abstract}


\section{Introduction}

In analyzing the structure of Minkowski space, Kopczy\'nski \&
Trautman \cite{KandT} stressed its underlying {\em affine} structure
which expresses the inertial properties of spacetime. Superimposed to
it, there exists a flat {\em metric} which allows to calculate
lengths, times, and angles. It is the combination of both structures,
the affine and the metric one, which determines the unique features of
the Minkowski space of special relativity.

In special relativity the effects of gravity are neglected. Then the
group of motions of spacetime is the Poincar\'e group with its $4+ 6$
parameters. This group is the semidirect product of the translation
and the Lorentz group.

The source of gravity is the {\em mass--energy} of matter. Because of
the Noether theorem, the {\em conservation} of mass--energy, in
Minkowski space, is related to the translation group. Consequently the
gauging of the translations should yield {\em gravity}, as foreseen by
Feynman \cite{Feynman}, amongst others. We understand here gauging as
a heuristic concept, as developed by Yang-Mills, which allows to
``deduce'' an interaction (here gravity) from a conserved current
(here mass-energy) and its associated invariance group (here the
translation group), see O'Raifeartaigh \cite{O'R}.

Since the translations are an inseparable part of the Poincar\'e
group, it is near at hand to gauge the Poincar\'e group altogether;
even more general, one can set free the ``frozen'' affine degrees of
freedom of the Minkowski space by gauging the affine group ${\rm
  A}(4,{\rm R})={\rm R}^4\semidirect {\rm GL}(4,{\rm R})$, the
semidirect product of the translations and the linear transformations.
If we still keep a metric superimposed to that affine gauge ansatz,
then we end up at the metric--affine gauge theory of gravity (MAG),
see the reviews \cite{PRs,Erice95} from which we also take our
conventions.

Andrzej Trautman \cite{Symposia,Tr3a} was one of the first to explore
the relation between the metric and the affine properties of
spacetime.  He restricted himself to the use of the curvature scalar
as a gravitational Lagrangian. Thus, he could not go beyond a
Riemann--Cartan spacetime with its metric compatible connection, see
theorem 3 on page 151 of ref.\cite{Symposia}. If one allows pieces in
the gravitational Lagrangian which are quadratic in the curvature, for
instance, then such a restriction becomes unnatural and one arrives at
the full potentialities of the metric--affine gauge theory of gravity.
We dedicate this article to Andrzej Trautman on the occasion of his
$2^{6\,th}$ birthday.

\section{A metric-affine model theory}

In metric--affine gravity, the metric $g_{\alpha\beta}$, the coframe
$\vartheta^{\alpha}$, and the connection $\Gamma_{\alpha}{}^{\beta}$
are considered to be independent gravitational field variables.  In
such a framework, one can recover general relativity by means of the
gravitational Lagrangian
\begin{equation}
  V_{\rm GR'}=-\frac{1}{2l^2}\left( R^{\alpha\beta}\wedge
    \eta_{\alpha\beta}+\beta Q\wedge\hspace{-0.8em}
    {\phantom{Q}}^{\star}Q\right)\,,
        \label{GR'}
\end{equation}
where we have $R_\alpha{}^\beta$ as curvature 2-form belonging to the
connection $\Gamma_{\alpha}{}^{\beta}$,
$\eta_{\alpha\beta}:=\hspace{-0.8em}
{\phantom{\theta}^{\star}}(\vartheta_{\alpha}\wedge
\vartheta_{\beta})$, the Weyl covector is $Q:=Q_{\gamma}^{\;\gamma}/4$
with the nonmetricity $Q_{\alpha\beta}:=-Dg_{\alpha\beta}$, and the
Hodge dual is denoted by a star $^\star$. Provided the matter Lagrangian
$L_{\rm mat}$ doesn't couple to the connection, i.e., if $\delta
L_{\rm mat}/\delta\Gamma_{\alpha}^{\;\beta}=0$, we fall back to
general relativity. It is decisive for these considerations to have a
non--vanishing $\beta$ in (\ref{GR'}), see \cite{HLS,Pono} for a
corresponding discussion.

In order to explore the potentialities of metric--affine gravity, we will
choose the simple non--trivial dilation--shear Lagrangian,
\begin{equation}
  V_{\rm dil-sh}=-\frac{1}{2l^2}\left( R^{\alpha\beta}\wedge
    \eta_{\alpha\beta}+\beta Q\wedge \hspace{-0.8em}
    {\phantom{Q}}^{\star}Q+ \gamma T\wedge \hspace{-0.8em}
    {\phantom{Q}}^{\star}T\right)-\frac{1}{8}\alpha\,
  R_{\alpha}{}^{\alpha}\wedge\hspace{-0.8em}
  {\phantom{Q}}^{\star}R_{\beta}{}^{\beta}\,,
         \label{dil-sh}
\end{equation}
with the dimensionless coupling constants $\alpha ,\;\beta$, and
$\gamma$, and $T:=e_{\alpha}\rfloor T^{\alpha}$, where $T^{\alpha}$ is
the torsion of spacetime. In future we will choose units such that
$l^2=1$. Observe that the last piece is of a pure {\em post}--Riemannian
nature, in fact, it is proportional to the square of Weyl's segmental
curvature which Weyl used in the context of his unsuccessful unified
field theory of 1918. It can be alternatively written as
$-(\alpha/2)\, {\rm d}Q \wedge\hspace{-0.8em}
{\phantom{Q}}^{\star}{\rm d}Q$, where $Q$ is the Weyl covector.

Below we will look for exact solutions of the field equations
belonging to the Lagrangian 
\begin{equation}\label{Ltot}L=V_{\rm dil-sh}+V_{\rm Max}\,,
  \qquad{\rm with}\qquad V_{\rm Max}=-(1/2)F\wedge\hspace{-0.8em}
  {\phantom{F}}^{\star}F
\end{equation} as the Lagrangian of the Maxwell field $F=dA$.  We will 
only be able to find non--trivial solutions, if the coupling constants
fulfill the constraint
\begin{equation}
  \gamma =-\frac{8}{3}\frac{\beta}{\beta +6}\;.
          \label{constraint}
\end{equation}
This is not completely satisfactory but, up to now, we cannot do
better. The search for exact solution in metric affine gravity has
been pioneered by Tresguerres \cite{Tres1,Tres2} and by Tucker and
Wang \cite{Tucker}.

\section{Starting with the Pleba\'nski-Demia\'nski\\ metric of general 
relativity}

Using the Eq.(3.30) of ref. \cite{pd}, see also
\cite{Alberto,blackhole}, the orthonormal coframe can be expressed in
terms of the coordinates $(\tau,q,p,\sigma)$ as follows,
\begin{eqnarray}
  \vartheta^{\hat{0}} &=&\frac{1}{H}\sqrt{\frac{{\cal Q}}{\Delta} }
  \left(d\tau - p^2 d\sigma \right)\,, \nonumber\\ \vartheta^{\hat{1}}
  &=&\frac{1}{H}\sqrt{\frac{\Delta}{{\cal Q}} }\; dq,\nonumber\\ 
  \vartheta^{\hat{2}} &=&\frac{1}{H}\sqrt{\frac{\Delta}{{\cal P}} }\;
  dp, \nonumber\\ \vartheta^{\hat{3}} &=&\frac{1}{H}\sqrt{\frac{{\cal
        P}}{\Delta} } \left(d\tau + q^2 d\sigma \right)\,, \label{U}
\end{eqnarray}
with the metric
\begin{equation} g=-\vartheta^{\hat{0}}\otimes \vartheta^{\hat{0}}
  +\vartheta^{\hat{1}}\otimes \vartheta^{\hat{1}}
  +\vartheta^{\hat{2}}\otimes \vartheta^{\hat{2}}
  +\vartheta^{\hat{3}}\otimes
  \vartheta^{\hat{3}}\,,\label{localmetric}\end{equation}
or
\begin{equation}
g= \frac{1}{H^2}\left\{- \frac{{\cal Q}}{\Delta}\;
(d \tau - p^2 d\sigma )^2 + \frac{\Delta}{{\cal Q}}\; dq^2 
 +\frac{\Delta}{{\cal P}}\; dp^2 +\frac{{\cal P}}{\Delta}\;
(d \tau + q^2 d \sigma)^2 \right\}\,.
\label{pd}
\end{equation}
The unknown functions are polynomials and read:
\begin{eqnarray}
  {\cal P} &:=& \left(\gamma - g_o^2- \frac{\lambda}{6}\right) +
  2n\, p - \epsilon \, p^2 + 2m \, p^3 - \left(\gamma + e_o^2 +
    \frac{\lambda}{6} \right) \, p^4\,, \nonumber\\ {\cal Q} &:=&
  \left(\gamma + e_o^2- \frac{\lambda}{6}\right) - 2m\, q +
  \epsilon \, q^2 - 2n \, q^3 - \left(\gamma - g_o^2 +
    \frac{\lambda}{6}\right) \, q^4\,, \nonumber\\ \Delta &:=& p^2 +
  q^2 \,,\nonumber \\ H &:=& 1 - p \, q\,.
\label{V}
\end{eqnarray}
The electromagnetic potential $A$ appropriate for this solution can be
expressed as follows:
\begin{eqnarray}
{ A}&=& \frac{1}{\Delta}\left[( e_o\,q + g_o\,p) { d \tau} +
         (g_o\,q - e_o\,p)\, p \,q { d \sigma} \right ]\nonumber\\
&=& \frac{H}{\sqrt{\Delta}}\left( \frac{e_o\, q }{\sqrt{{\cal
          Q}}} \;\vartheta^{\hat{0}} + \frac{g_o\,
      p}{\sqrt{{\cal P}}} \;\vartheta^{\hat{3}}\right).
\label{W}
\end{eqnarray}

\section{Generating solutions in metric-affine gravity}

It has been pointed out by Dereli, Tucker, et al., see
\cite{TuckO,TuckerJadwisin}, and by Obukhov et al., see \cite{OVEH},
that exact solutions of metric-affine gravity can be generated from
electrovacuum solutions of general relativity if one assumes, besides
the coframe (\ref{U}), (\ref{localmetric}), a fairly simple form for
nonmetricity and torsion, namely the so-called triplet ansatz (see
\cite{OVETH}) patterned after the electromagnetic potential in
(\ref{W}):
\begin{eqnarray}
  Q &=& k_0\frac{H}{\sqrt{\Delta}}\left( \frac{ N_{\rm e}\, q }{\sqrt{{\cal
          Q}}} \;\vartheta^{\hat{0}} + \frac{N_{\rm g}\,
      p}{\sqrt{{\cal P}}} \;\vartheta^{\hat{3}}\right),\nonumber \\ T
  &=& k_1\frac{H}{\sqrt{\Delta}}\left( \frac{ N_{\rm e}\, q }{\sqrt{{\cal
          Q}}}\; \vartheta^{\hat{0}} + \frac{N_{\rm g}\,
      p}{\sqrt{{\cal P}}}\; \vartheta^{\hat{3}}\right),
\label{triplet}\\
\Lambda&=& k_2\frac{H}{\sqrt{\Delta}}\left( \frac{N_{\rm e}\,
    q}{\sqrt{{\cal Q}}} \; \vartheta^{\hat{0}} + \frac{N_{\rm g}\,
    p}{\sqrt{{\cal P}}}\; \vartheta^{\hat{3}}\right).  \nonumber
\end{eqnarray}
The Weyl covector $Q$ and the torsion trace $T$ were defined above,
the $\Lambda$ co\-vector represents one irreducible piece of the shear
part of the nonmetricity $\Lambda:=\vartheta^\alpha
e^\beta\rfloor(Q_{\alpha\beta}- Qg_{\alpha\beta})$. \medskip

The result of ref.\cite{OVEH} is even stronger: Let the starting point
be a general metric-affine Lagrangian which is quadratic in curvature,
torsion, and nonmetricity. Provided this Lagrangian possesses, like
(\ref{dil-sh}), only one curvature {square} piece built up from Weyl's
{\em segmental} curvature $R_\alpha{}^\alpha$, then the triplet
(\ref{triplet}) represents the general solution for the post-Riemannian
pieces of the connection. In other words, if one wants to find exact
solutions encompassing also the other $2+2$ irreducible pieces of
nonmetricity and torsion, respectively, then one has to enrich the
Lagrangian by other curvature square pieces.

For our Lagrangian (\ref{dil-sh}), the constants in the triplet
(\ref{triplet}) read
\begin{equation}\label{k0k1k2}
  k_0=-\frac{24}{\beta+6}\,,\quad k_1=-\frac{36}{\beta+6}\,,\qquad k_2=6\,,
\end{equation}
and $N_{\rm e}$ and $N_{\rm g}$ are the quasi-electric and the
quasi-magnetic dilation--shear--spin charges, respectively. Now, in
MAG, the polynomials depend also on these quasi-charges in a fairly
trivial way,
\begin{eqnarray}
  {\cal P} &:=& \left(b - g_o^2 - {\cal G}_o^2\right) +
  2np - \epsilon p^2 + 2 m \mu p^3 - \left[\mu^2 \left(b +
      e_o^2+{\cal E}_o^2 \right) + \frac{\lambda}{3}
  \right] p^4 ,\nonumber\\ {\cal Q} &:=& \left(b + e_o^2
      +{\cal E}_o^2\right) - 2mq + \epsilon q^2 - 2n
  \mu q^3 - \left[ \mu^2 \left(b -g_o^2-{\cal G}_o^2
      \right) + \frac{\lambda}{3} \right] q^4 ,\nonumber \\ 
  \Delta &:=& p^2 + q^2 ,\nonumber \\ H &:=& 1 - \mu \, p \, q\, ,
\label{solution-n}
\end{eqnarray}
where we introduced $b:=\gamma - {\lambda}/{6}$. We slightly
generalized the solution by means of the parameter $\mu$, which can
take the values $-1,0,+1$. The post-Riemannian charges $N_{\rm e}$ and
$N_{\rm g}$ are related to the post-Riemannian pieces ${\cal E}_o$ and
${\cal G}_o$, entering the polynomials, according to
\begin{equation}\label{constraint'}
  {\cal E}_o=k_0\sqrt{\frac{\alpha}{2}}\,N_{\rm e}\,,\qquad {\cal G}_o
  =k_0\sqrt{\frac{\alpha}{2}}\,N_{\rm g}\,,\qquad{\rm with}\qquad
  \alpha >0\,.
\end{equation}

The solution (\ref{U}), (\ref{localmetric}), (\ref{triplet}) to
(\ref{constraint'}) has been thoroughly checked with the help of our
computer algebra programs written in Reduce for the exterior calculus
package Excalc, see \cite{Stauffer}. This solution seems to exhaust
all the possibilities one has with the Pleba{\'n}ski-Demia\'nski
metric and the triplet ansatz.

Finally, we would like to link up our new solution with the more
special cases known from the literature \cite{OVETH,VTOH,PLH,eo=0}.

\section{Reduction to a solution with mass, angular momentum, 
electric charge and quasi-electric post-Riemannian triplet}

In order to recover known solutions, we change to more familiar
coordinates, namely to the Boyer-Lindquist coordinates of the Kerr
solution:
\begin{equation}\label{trafo}
  (\tau, q,p, \sigma) \longrightarrow (t,y,x,\phi)\longrightarrow
  (t,r,\theta,\phi)\,.\end{equation} More exactly, we have
\begin{eqnarray}
 \tau &=& j_o (t - j_o \phi)\,,\nonumber \\
\sigma &=& - j_o^2\, \phi\,, \nonumber \\
q &=& \frac{y}{j_o}= \frac{r}{j_o}\,,\nonumber \\
p &=& x =-\cos \theta\,.\label{trafo1}
\end{eqnarray}
By means of these transformations, the coframe (\ref{U}) has the same
form as that of the Vtoh-solution of ref.\cite{VTOH}, Eq.(3.1). In
fact, we can identify the coframes, provided we have 
\begin{equation}
\frac{1}{H}\sqrt{\frac{{\cal Q} j_o^2}{\Delta}}\equiv\sqrt{
\frac{\Delta_{\rm Vtoh}}{\Sigma_{\rm Vtoh}}}\; ,\qquad
\frac{1}{H}\sqrt{\frac{{\cal P}}{\Delta}}\equiv\sqrt{
\frac{f_{\rm Vtoh}\sin^2\theta}{\Sigma_{\rm Vtoh}}}\;.
\end{equation}
These identities can be fulfilled by the ansatz
\begin{equation}
  H=1\,,\quad\Delta j_o^2=\Sigma_{\rm Vtoh}\,,\quad {\cal
    Q}j_o^4=\Delta_{\rm Vtoh}\,,\quad{\cal P}j_o^2=f_{\rm
  Vtoh}\sin^2\theta\,.\label{ident}
\end{equation}
Therefore we have first to kill the $\mu$--parameter: $\mu=0$.
Furthermore, the magnetic and the quasi--magnetic charges and the
NUT--parameter have to vanish:
\begin{equation}
  g_o=0\,,\quad{\cal G}_o=0\,,\quad N_{\rm g}=0\,,\quad
  n=0\,.\label{mag}\end{equation} A modification of $\epsilon
=1-{\lambda}j_o^2/{3}$ is also necessary.\medskip

Then, by a suitable redefinition of the constants, the electromagnetic
potential reduces to (cf.\cite{ZiF})
\begin{equation}
  { A}= \frac{e_o\,r}{r^2+j_o^2\cos^2\!\theta}\left(d t -
    j_o\sin^2\!\theta d \phi\right)\,,
\end{equation}
and the Vtoh-functions (for $a_0=1$ and $z_4=\alpha$) turn out to be:
\begin{eqnarray}
  \Delta_{\rm Vtoh}&=&r^2+j_o^2-2 M
  r-\frac{\lambda}{3}r^2\left(r^2+j_o^2 \right) +\alpha\frac{ (k_0
    N_{\rm e})^2}{2} + e_o^2\,,\nonumber\\ \Sigma_{\rm Vtoh}&=&r^2+j_o^2
  \cos^2\!\theta\,, \nonumber\\ f_{\rm
    Vtoh}&=&1+\frac{\lambda}{3}j_o^2\cos^2\!\theta\,.
\end{eqnarray}
Note that we kept the electric charge $e_o$. Correspondingly, we found
the charged version of the Vtoh-solution. Putting $e_o=0$, we finally
recover the Vtoh-solution of \cite{VTOH}: It carries only mass,
angular momentum, and the quasi--electric dilation--shear--spin charge
$N_{\rm e}$. Our solution of Sec.4 has, additionally, the
NUT--parameter, electric and magnetic charges, the acceleration
parameter, and the quasi--magnetic dilation--shear--spin charge.

\section{\bf Acknowledgments}

\noindent We would like to thank Alberto Garc\'{\i}a, Alfredo 
Mac\'{\i}as, and Yuri Obukhov for interesting discussions and remarks.
This research was supported by CONACyT-M\'exico, grant No.\ 
3898P--E9608 and by the joint German--Mexi\-can project Conacyt--DLR
E130--2924 and DLR--Conacyt MXI 6 B0A 6A. Moreover, J.S.\ acknowledges
support from the ANUIES--DAAD agreement, Kennziffer A/98/04459.

\end{document}